\documentstyle[12pt]{article}

\def\e{{\rm e}}
\newcommand{\be}{\begin{equation}}
\newcommand{\ee}{\end{equation}}
\newcommand{\bea}{\begin{eqnarray}}
\newcommand{\eea}{\end{eqnarray}}
\newcommand{\nn}{\nonumber}
\begin{document}
\parindent=1.5pc
\thispagestyle{empty}
\rightline{hep-th/9409185}
\rightline{MPI-PhT/94-64}
\rightline{September 1994}
 \begin{center}{
{\bf A REMARK ON THE INSTABILITY OF \\
              \vglue 10pt
THE BARTNIK-McKINNON SOLUTIONS
               }
\vglue 5pt
\vglue 1.0cm
{GEORGE LAVRELASHVILI
 \footnote{On leave of absence from Tbilisi
Mathematical Institute, 380093 Tbilisi, Georgia}
 }\\
\baselineskip=14pt
{\it Max-Planck-Institut f\"ur Physik, Werner-Heisenberg-Institut}\\
{\it F\"ohringer  Ring 6, 80805 Munich, Germany}\\
\vglue 0.3cm
{and}\\
\vglue 0.3cm
{DIETER MAISON}\\
{\it Max-Planck-Institut f\"ur Physik, Werner-Heisenberg-Institut}\\
{\it F\"ohringer Ring 6, 80805 Munich, Germany}\\
\vglue 0.8cm
{ABSTRACT}
}\end{center}
\vglue 0.3cm
{\rightskip=3pc  \leftskip=3pc
\tenrm\baselineskip=12pt
\noindent
The aim of the present letter is to critically review the
stability of the Bartnik-McKinnon
solutions of the Einstein-Yang-Mills theory.
The stability question was already studied by several
authors, but there
seems to be some confusion about the nature and the number
of unstable modes.
We suggest to distinguish two different kind of instabilities,
which we call `gravitational' respectively `sphaleron' instabilities.
We claim that the $n^{\rm th}$
Bartnik-McKinnon solution has exactly
$2n$ unstable modes, $n$ of either type.

\vglue 0.8cm}
\newpage
\noindent
The discovery of a discrete family of
globally regular, static spherically symmetric
solutions of Einstein-Yang-Mills (EYM) theory
by Bartnik and McKinnon \cite{BM} gives rise to the question
of their physical interpretation.

It was argued \cite{VG,SW} that they are gravitational
analogues of the electroweak sphale\-rons \cite{SPH}.
One of the key points of this interpretation is the presence
of unstable mode(s)
in the fluctuation spectrum around these solutions
\cite{SZ,Z,M,BROD,VG2}.
Since the solutions are based on a very special ansatz for the
fields one may study as well fluctuations within this ansatz
as more general ones.
An analysis of fluctuations of the first type was performed in
\cite{SZ,Z,M}, with the result that
$n$ negative modes were found for the $n^{\rm th}$
Bartnik-McKinnon (BMK) solution.
Inspired by this `empirical' observation Sudarsky and Wald
\cite{SW} forwarded an
argument that there ought to be just that number of unstable
modes of the solutions due to their nature as `sphalerons'
interpolating between topologically inequivalent vacua.
In addition they suggested the existence of an additional instability
with respect to a (nonlinear) rescaling of the solutions.
Thus they claimed that the $n^{\rm th}$ BMK solution has in fact
$n+1$ unstable directions. This number of negative modes
was `confirmed' in ref. \cite{VOL}.
Furthermore an analytical proof of the instability of the BMK
solutions under spherically symmetric perturbations of the
second kind was given in ref. \cite{BROD,VG2}.

The aim of the present letter is to critically review these results
and clarify some confusing points. In fact, our analysis reveals that
the $n^{\rm th}$ solution has $n$ unstable modes of the first type, which
may be called `gravitational' instabilities since they have no analogue
for the flat-space sphalerons of the YM-Higgs theory.
These are the instabilities discussed in \cite{SZ,Z,M}.
At least for the lowest
member of the BMK family this instability is related to the scaling
instability considered  by Sudarsky and Wald. In addition there are
$n$ unstable modes of the second type, which may be called
`sphaleron' instabilities, because they correspond to the unstable mode
found for the YM-Higgs sphaleron \cite{YA}. This type of instability
of the BMK solutions was discussed in \cite{VG,BROD,VG2}.
Altogether we find that the
$n^{\rm th}$ BMK solution has $2n$ negative modes within the most general
spherically symmetric ansatz.

For our analysis we will essentially adopt the
notations of ref. \cite{BFM}.
Due to the spherical symmetry of the solutions the space-time manifold
splits into a product $M_2\times S^2$. The dynamics of the theory
can be `dimensionally reduced' to a 2-dimensional theory on $M_2$.
The line element
decomposes as $ds^2=ds^2_2+r^2d\Omega^2$, where $d\Omega^2$ is the
invariant line element of the unit 2-sphere. For the metric on $M_2$
we choose the parametrization
\be
ds^2_2=A^2(t,r)\mu(t,r)dt^2-{dr^2\over\mu(t,r)}\;.
\label{metric}
\ee
The most general static, spherically symmetric
ansatz for the {\sl SU(2)\/} Yang-Mills field $W_\mu^a$ can be written
(in the Abelian gauge) as
\bea
 W_t^a&=(0,0,A_0)\,,\qquad  W_\theta^a&=(\phi_1,\phi_2,0)\, \nn \\
 W_r^a&=(0,0,A_1)\,, \qquad
 W_\varphi^a&=(-\phi_2 \sin\theta,\phi_1\sin\theta,\cos\theta)\,.
\label{gauge}
\eea
This ansatz (\ref{gauge}) is form invariant
under gauge transformations around the third isoaxis,
with $A_\alpha$ transforming as a $U(1)$ gauge field
on the reduced space-time $M_2$, whereas
$\phi=\phi_1+i\phi_2$ is a scalar of charge one with the
covariant derivative
$D_\alpha\phi=\partial_\alpha\phi-iA_\alpha\phi$.
With respect to this $U(1)$ one may define the `charge conjugation'
$\phi\to\overline\phi$, $A_\alpha\to -A_\alpha$.
The reduced EYM action is $S=S_G+S_{\it YM}$ with
($F_{\alpha\beta}$ denotes the field strength of $A_\alpha$)
\be
S_G=-{1\over 2}\int drdtA(\mu+r\mu'-1)
\label{actG}
\ee
and
\be
S_{\it YM}=-\int drdtA\Bigl[{r^2\over4}F^{\alpha\beta}F_{\alpha\beta}
-\overline{D^\alpha\phi} D_\alpha\phi+
{1\over2r^2}(|\phi|^2-1)^2\Bigr]\;.
\label{actYM}
\ee
in units where the Newton constant $G$ and the gauge coupling $g$ are
set to one.
Choosing the gauge $A_0=0$ the action $S_{\it YM}$ can
be written more explicitly
\be
S_{\it YM}=-\int drdtA\Bigl[-{r^2\over2A^2}\dot A_1^2-
{1\over A^2\mu}|\dot\phi|^2
           +\mu|\phi'-iA_1\phi|^2+{1\over2r^2}(|\phi|^2-1)^2\Bigr]\;,
\label{act1}
\ee
(where a prime denotes $d\over dr$ and a dot $d\over dt$).

Bartnik and McKinnon \cite{BM}
found numerically the first few members of an
infinite family of static solutions of the field equations derived
from $S_G+S_{\it YM}$. The existence of this family was later
proved analytically in refs. \cite{SMO,BFM}.
These solutions are even under the $U(1)$
charge conjugation ($\phi_2=A_1=0$) and
can be labelled by the number of zeros of the only remaining
component $\phi_1\equiv W$ of the YM field.
In order to analyse their stability under small perturbations
we have to consider the spectrum of
(harmonically) time dependent perturbations in the background
of the BMK solutions.
The existence of solutions of the linearized field equations
corresponding to imaginary frequency
indicates the instability of the background solution leading
to an exponential growth of the perturbation in time.

For perturbations of the type
\bea
\phi_1\to W(r)+\varphi(r)\e^{i\omega t}&\quad&
A\to A(r)+a(r) e^{i\omega t}\cr
\phi_2\to \psi(r)\e^{i\omega t}&\quad&
\mu \to \mu(r)+\kappa(r) e^{i\omega t}\cr
A_1\to a_1e^{i\omega t}&\quad&
\eea
the linearized field equations are
\bea
A\mu\Bigl(W'\psi-W(\psi'-a_1W)\Bigr)&=&{\omega^2r^2\over2A}a_1\cr
-(A\mu\varphi')'-\Bigl((A\kappa+\mu a)W'\Bigr)'+A{3W^2-1\over r^2}\varphi
+{W(W^2-1)\over r^2}a&=&{\omega^2\over{A\mu}}\varphi\cr
-\Bigl(A\mu(\psi'-a_1W)\Bigr)'+A\mu W'a_1+A{W^2-1\over r^2}\psi&=&
{\omega^2\over{A\mu}}\psi\cr
ra'-4AW'\varphi'-2W'^2a&=&0\cr
(r\kappa)'+2W'^2\kappa+4\mu W'\varphi'+{4W(W^2-1)\over r^2}\varphi&=&0\;,
\label{lineq}
\eea
with $W$, $\mu$ and $A$ the background solutions.
The last equation can be integrated imposing the boundary
condition $\kappa (\infty)=0$ with the result
\be
\kappa =-{4\mu W'\over r}\varphi\;.
\label{kappa}
\ee
Putting back this expression for $\kappa$ into the equation for
$\varphi$ allows to
eliminate the gravitational degrees of freedom
$\kappa$ and $a$ from the equations for the YM field.
Furthermore the equations for the even and odd sectors under the
$U(1)$ charge conjugation decouple. One obtains
\be
-(A\mu\varphi')'+A{3W^2-1\over r^2}\varphi
+\Bigl({4A\mu W'^2\over r}\Bigr)'\varphi=
{\omega^2\over{A\mu}}\varphi,
\ee
and
\bea
A\mu\Bigl(W'\psi-W(\psi'-a_1W)\Bigr)&=&{\omega^2r^2\over2A}a_1\cr
-\Bigl(A\mu(\psi'-a_1W)\Bigr)'+A\mu W'a_1+A{W^2-1\over r^2}\psi&=
&{\omega^2\over A\mu}\psi\;.
\eea
The $\varphi$-sector was analyzed in the earlier works
\cite{SZ,Z,M} with the result that $n$
negative modes for the $n^{\rm th}$ BMK solution were found.
Up to now this coincidence was considered as a curiosity for
which no explanation was offered. Although we also cannot prove
this coincidence, we want to give some `explanation' for it.
One may add a mass term
\be
S_m=-{\alpha^2\over4}\int drdtA\Bigl((\phi_1+1)^2+\phi_2^2\Bigr)
\ee
to the action for the YM field.
Such a term results from the coupling of a Higgs
field (doublet) in the limit of infinite Higgs mass. The resulting
theory was studied in \cite{GMN}. As the mass $\alpha$ is varied
one finds one-parameter families of solutions tending to the
BMK solutions in the limit $\alpha\to 0$. These families have the
property that they interpolate continuously between the $n^{\rm th}$ and
$n+1^{\rm st}$ BMK solution
as $\alpha$ increases from zero to some maximal value
$\alpha^{(n)}_{\rm max}$ and then runs back to zero.
The points $\alpha=\alpha^{(n)}_{\rm max}$ are bifurcation points at
which the branch starting from the
$n^{\rm th}$ BMK solution
and the one
starting from the $n+1^{\rm st}$ BMK solution merge.
It is well known that at such bifurcation points the number of
unstable modes changes generically by one.
Thus starting with the stable trivial solution
(Minkowski space, $W\equiv 1$) we end up with one
unstable mode for the first non-trivial BMK solution and so on.

A further remark may be at order. Although it is true that
for the lowest BMK solution the scaling of the solution
proposed in \cite{SW} yields an unstable direction this is not
generally so for the higher ($n>2$) BMK solutions \cite{M}.

Next let us turn to the $U(1)$ charge conjugation odd $\psi$ sector.
As already mentioned an analytical proof of the existence of at least
one unstable mode  of this type has been given in \cite{BROD,VG2}.
We have performed a numerical analysis  of the fluctuation spectrum
for the first few members of the BMK family.
We find that the $n^{\rm th}$
solution has $n$ negative modes also in the $\psi$ channel.
In fact, this is what was suggested by Sudarsky and Wald \cite{SW},
ignoring that they interpreted these instabilities as the
ones found in the $\varphi$ sector.

Hence altogether we claim that the
$n^{\rm th}$ BMK solution has $2n$ unstable modes with respect to
spherically symmetric perturbations.

For comparison we have collected in Tables 1 and 2
our numerical values
for the energies $E=\omega^2$ of the negative modes of
the first three BMK solutions in the $\varphi$ resp.\ $\psi$ sectors.

\begin{center}
\begin{tabular}{|l|l|l|}  \hline
$N=1$        &$N=2$        &$N=3$          \\ \hline
$E_1=-0.0525$&$E_1=-0.0410 $&$E_1=-0.0339 $ \\
             &$E_2=-0.0078 $&$E_2=-0.0045 $ \\
             &              &$E_3=-0.0006 $ \\  \hline
\end{tabular}
\vglue 0.4cm
Tab 1. Boundstate energies for the $N=1,2,3$  BMK solutions,
($\varphi$ sector).
\end{center}

\begin{center}
\begin{tabular}{|l|l|l|l|}  \hline
$N=1$        &$N=2$        &$N=3$         \\ \hline
$E_1=-0.0619$&$E_1=-0.0360$&$E_1=-0.0346$ \\
             &$E_2=-0.0105$&$E_2=-0.0037$ \\
             &             &$E_3=-0.0009$ \\ \hline
\end{tabular}
\vglue 0.4cm
Tab 2. Boundstate energies for the $N=1,2,3$  BMK solutions,
($\psi$ sector).
\end{center}


\begin{thebibliography}{9}
\bibitem{BM}
 R. Bartnik and J. McKinnon, {\it Phys. Rev. Lett.}
{\bf 61} (1988) 141.

\bibitem{VG}
 M.S. Volkov and D.V. Gal'tsov, {\it  Phys. Lett.}
{\bf B273} (1991) 255.

\bibitem{SW}
D. Sudarsky and R.M. Wald, {\it Phys. Rev.}
{\bf D46} (1992) 1453.

\bibitem{SPH}
R.F. Dashen, B. Hasslacher, and A. Neveu, {\it Phys. Rev.}
{\bf D10} (1974) 4138;\\
N. Manton, {\it Phys. Rev.}
{\bf D28} (1983) 2019;\\
F.R. Klinkhamer and N.S. Manton, {\it Phys. Rev.}
{\bf D30} (1984) 2212.

\bibitem{SZ}
N. Straumann and Z.H. Zhou, {\it Phys. Lett.}
{\bf B243} (1990) 33.

\bibitem{Z}
Z.H. Zhou, {\it Helv. Phys. Acta}
{\bf 65} (1992) 767.

\bibitem{M}
 D. Maison, (1989), {\it unpublished.}

\bibitem{BROD}
O. Brodbeck and N. Straumann, {\it Phys. Lett.}
{\bf B324} (1994) 309.

\bibitem{VG2}
 M.S. Volkov and D.V. Gal'tsov, {\it Odd-parity negative modes of
 Einstein-Yang-Mills black holes and sphalerons,}
 Preprint ZU-TH 27/94, hep-th/9409041, 1994.

\bibitem{VOL}
 M.S. Volkov, {\it  Phys. Lett.}
{\bf B328} (1994) 89.

\bibitem{YA}
J. Burzlaff, {\it  Nucl. Phys.}
{\bf B233} (1984) 262;\\
 L. G. Yaffe, {\it Phys. Rev.}
{\bf D40} (1989) 3463.

\bibitem{BFM}
 P. Breitenlohner, P. Forg\'acs and D. Maison,
 {\it Commun. Math. Phys.}
{\bf 163} (1994) 141.

\bibitem{SMO}
J. A. Smoller and A.G. Wasserman, {\it Commun. Math. Phys.}
{\bf 151} (1993) 303.

\bibitem{GMN}
 B.R. Greene, S.D. Mathur and C.M. O'Neill,
 {\it Phys. Rev.}
 {\bf D47} (1994) 2242.

%
\end{thebibliography}
\end{document}